\documentclass[]{WileyASNA-v1}

\usepackage[utf8]{inputenc}
\usepackage{calc}
\usepackage{hyperref}
\usepackage{endnotes}

\newlength{\okinalen}
\newcommand{\okina}{\hbox to.666\okinalen{\hss`\hss}}

\articletype{Article Type}

\received{2022}
\revised{2022}
\accepted{2022}

\begin{document}

\title{Follow-up observations of the binary system $\gamma$\,Cep}

\author[1]{M. Mugrauer}

\author[1]{S. Schlagenhauf}

\author[2,3]{S. Buder}

\author[4]{C. Ginski}

\author[5]{M. Fern\'{a}ndez}

\authormark{Mugrauer et al.}

\address[1]{Astrophysikalisches Institut und Universit\"{a}ts-Sternwarte Jena}

\address[2]{Research School of Astronomy \& Astrophysics, Australian National University, Canberra, ACT 2611, Australia}

\address[3]{ARC Centre of Excellence for All Sky Astrophysics in 3 Dimensions (ASTRO 3D), Australia}

\address[4]{Leiden Observatory, Leiden University, PO Box 9513, 2300 RA Leiden, The Netherlands}

\address[5]{Instituto de Astrof\'isica de Andaluc\'ia CSIC, Glorieta de la Astronomia, 3 18008 Granada (SPAIN)}

\corres{M. Mugrauer, Astrophysikalisches Institut und Universit\"{a}ts-Sternwarte Jena, Schillerg\"{a}{\ss}chen 2, D-07745 Jena, Germany.\newline \email{markus@astro.uni-jena.de} \thanks{Data here reported were acquired at Centro Astron\'omico Hispano Alem\'an (CAHA) at Calar Alto operated jointly by Instituto de Astrof\'isica de Andaluc\'ia (CSIC) and Max Planck Institut f\"{u}r Astronomie (MPG). Centro Astron\'omico Hispano en Andaluc\'{\i}a is now operated by Instituto de Astrof\'isica de Andaluc\'ia and Junta de Andaluc\'ia.}}

\abstract{We present follow-up imaging and spectroscopic observations of the exoplanet host star $\gamma$\,Cep\,A, and of its low-mass stellar companion $\gamma$\,Cep\,B. We used the lucky-imager AstraLux at the Calar Alto observatory to follow the orbital motion of the companion around its primary, whose radial velocity was determined with spectra of the star, taken with the \'Echelle spectrograph FLECHAS at the University Observatory Jena. We measured the astrometry of the companion relative to the exoplanet host star in all AstraLux images and determined its apparent SDSS i$'$-band photometry, for which we obtained $\text{i}'=9.84\pm0.17$\,mag. Using stellar evolutionary models and the \textit{Gaia} parallax of the exoplanet host star, we derived the mass of $\gamma$\,Cep\,B to be $0.39\pm0.03$\,M${_\odot}$. This is in good agreement with the mass of the companion that we derived from its near-infrared photometry given in the literature. With the detection limit reached in our AstraLux images we explored the detection space of potential additional companions in the $\gamma$\,Cep binary system. In the background limited region at angular separations larger then 5\,arsec (or 69\,au of projected separation) companions down to $0.11$\,M$_\odot$ are detectable around the bright exoplanet host star. The radial field of view, fully covered by our AstraLux observations, exhibits a radius of 11.2\,acsec. This allows the detection of companions with projected separations up to 155\,au. However, except for $\gamma$\,Cep\,B no additional companions could be imaged with AstraLux around the exoplanet host star. We redetermined the orbital solution of the $\gamma$\,Cep binary system with the new AstraLux astrometry of $\gamma$\,Cep\,B and the additional radial velocities of $\gamma$\,Cep\,A, obtained from our FLECHAS spectroscopy of the star, combined with astrometric and radial velocity data from the literature. The determined Keplerian orbital elements were used to derive the system parameters and to calculate specific future ephemeris for this intriguing exoplanet host binary star system.}


\maketitle

\section{Introduction}

The third brightest star in the northern constellation Cepheus, $\gamma$\,Cep (alias HD\,222404, or HIP\,116727), is a nearby (sub)giant, which is visible to the naked eye \citep[$G \sim 2.9$\,mag, $D \sim 13.8$\,pc, $SpT=\text{K1III-IV}$,][]{gaiaedr3, Keenan1989}. With precise radial velocity (RV) measurements \cite{hatzes2003} could detect a jovian exoplanet, which revolves around this star on a 2.5 year orbit ($a \sim2 $\,au), and has a minimum-mass, which is about 1.7 times the mass of Jupiter in our solar system. Furthermore, the exoplanet host star $\gamma$\,Cep itself was identified to be the primary component of a binary system. The stellar multiplicity of the star was first reported by \cite{campbell1988}, who found a decrease of its RV of about 0.26\,km\,s$^{-1}$\,yr$^{-1}$ within a range of time of 6\,yr, indicating that $\gamma$\,Cep is a single-lined spectroscopic binary with an orbital period of several decades. The first spectroscopic orbital solution for this binary system was then derived by \cite{griffin2002} with RV measurements, which were taken in a span of time of more than 100\,yr. \cite{torres2007} constrained this orbital solution by using beside the RV data from \cite{griffin2002} also additional observational data of the star, which he compiled from the literature, as well as new RV measurements of $\gamma$\,Cep, which he derived from available archival spectra. Eventually, \cite{neuhaeuser2007} could directly detect the secondary $\gamma$\,Cep\,B next to the exoplanet host star using adaptive optics and seeing limited imaging in the near-infrared. In three observing epochs in 2006, the angular separation ($\rho$) and the position angle ($PA$) of the stellar companion relative to its primary could be measured, which allowed a further refinement of the orbital solution of the $\gamma$\,Cep binary system and of its parameters (semi-major axis and dynamical masses of its components). Furthermore, in the taken images the infrared photometry of $\gamma$\,Cep\,B was determined, which can be used for an independent mass estimation of the companion with stellar evolutionary models (see section 2).

According to the orbital solution, derived by \cite{neuhaeuser2007}, the $\gamma$\,Cep system is an eccentric ($e=0.4112\pm0.0063$) binary with a semi-major axis of $20.18\pm0.66$\,au, which makes it to one of the closest known exoplanet host multiple star systems with an exoplanet orbiting one of its components. As listed in the \textsl{Extrasolar Planets Encyclopaedia}\footnote{\url{http://exoplanet.eu/planets_binary/}} \citep{schneider2011}, currently only about a dozen of these systems ($a \lesssim 20$\,au) are known. Hence, $\gamma$\,Cep is an intriguing system to test models for planet formation in close binaries and models that characterize the long-term orbital stability of planets in these systems. In a case like that of $\gamma$\,Cep the planet-bearing disk around the primary star gets truncated at about 5\,au by the gravitational impact of the secondary star \citep{jangcondell2008}, which also can significantly influence the disk properties, e.g. its morphology, temperature \& density distribution, or the relative velocity of colliding planet forming planetesimals \citep[see e.g.][]{mayer2005, marzari2000}. Hence, the planet formation in this system is limited to the vicinity of the exoplanet host star, which also holds for the region of long-term stable orbits around the star \citep[$a_\text{planet} < a_\text{critical}$, defined by][with $a_\text{critical}=3.85\pm0.38$\,au, as determined by Neuh\"auser et al. 2007]{holman1999}.

According to the orbital solution of the $\gamma$\,Cep binary system from \cite{neuhaeuser2007}, after 2013.5 the secondary should be separated from the bright exoplanet host star by more than 1.4\,arcsec, which makes it detectable also for telescopes of the 2\,m-class using high contrast or Seeing enhanced imaging techniques. In order to further follow the orbital motion of $\gamma$\,Cep\,B around its primary we have added the star to the target list of our multiplicity study of exoplanet host stars \citep[whose first results are described by][]{ginski2012, ginski2016}, carried out with the lucky-imager AstraLux \citep{hormuth2008}, which is operated at the Cassegrain focus of the 2.2\,m telescope of the Calar Alto Observatory in Spain. In addition, we have obtained additional RV measurements of $\gamma$\,Cep in the course of the Gro{\ss}schwabhausen (GSH) Binary Survey \citep[see e.g.][]{mugrauer2017,bischoff2017} using the \'Echelle spectrograph FLECHAS \citep{mugrauer2014}, which is installed at the Nasmyth focus of the 0.9\,m telescope of the University Observatory Jena. In the following two sections of this paper we describe our AstraLux and FLECHAS observations, and present new astrometric measurements of $\gamma$\,Cep\,B, as well as RVs of the exoplanet host star. These measurements combined with additional observational data, compiled from the literature, were used to determine a new orbital solution of the $\gamma$\,Cep binary system, which is presented in the final section of this paper.

\section{AstraLux Observations}

In the course of our multiplicity study of northern exoplanet host stars, carried out at the Calar Alto Observatory in Spain, $\gamma$\,Cep was observed in the SDSS i$'$-band \citep{fukugita1996} with AstraLux in 5 nights between 2014 and 2020. Details of the observations are summarized in the AstraLux observing log, which is shown in Table\,\ref{TAB_ASTRALUX_LOGFILE}\hspace{-1.75mm}.

\begin{table}[h!]
\caption{AstraLux observing log. For each observing date ($ObsDate$) the number of frames ($N_{\rm Frames}$), the used detector integration time ($DIT$), the resulting total integration time ($TIT$) on target, as well as the average airmass ($X$) and atmospheric Seeing during the observations are listed.}\label{TAB_ASTRALUX_LOGFILE}
\centering
\begin{tabular}{ccccccc}
\hline
$ObsDate$   & $N_{\rm Frames}$ & $DIT$    & $TIT$   & $X$  & Seeing\\
             &              & $[$ms$]$ & $[$s$]$ &      & $[$arcsec$]$\\
\hline
21. Aug 2014 & 20000        & 15.01    & 300     & 1.33 & 1.1         \\
27. Aug 2015 & 25000        & 12.51    & 313     & 1.32 & 1.1         \\
26. Oct 2016 & 10000        & 29.54    & 295     & 1.32 & 1.0         \\
30. Oct 2017 & 25000        & 29.54    & 739     & 1.41 & 1.5         \\
05. Sep 2020 & 50000        & 30.00    & 1500    & 1.32 & 0.8         \\
\hline
\end{tabular}
\end{table}

\begin{figure*}
\resizebox{\hsize}{!}{\includegraphics{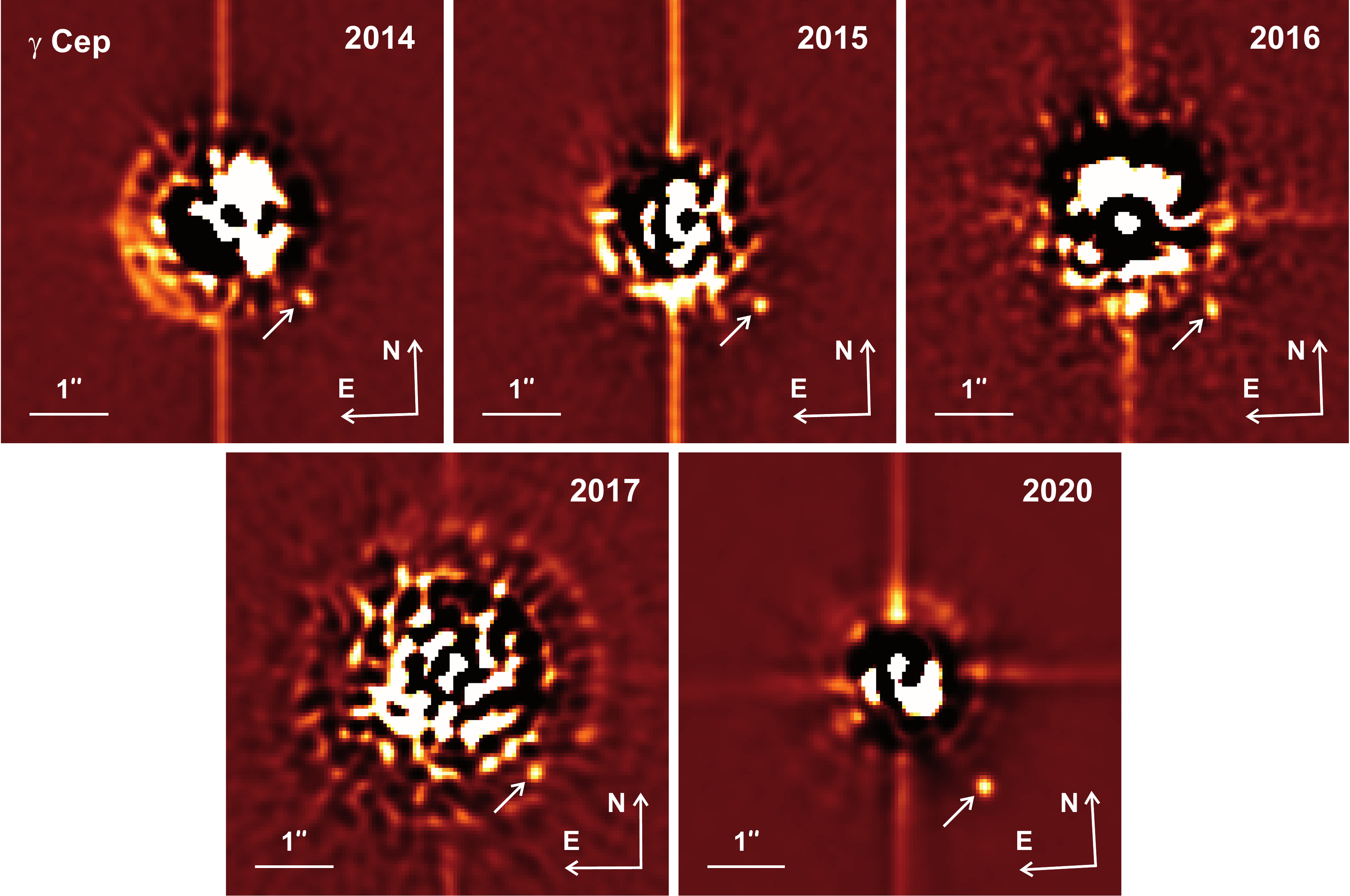}}\caption{Detail views of our fully reduced AstraLux i$'$-band images, which show a field of view of 5.6\,arcsec $\times$ 5.6\,arcsec, centered on the exoplanet host star $\gamma$\,Cep\,A. The stellar companion $\gamma$\,Cep\,B, located southwest of the star, is marked with a white arrow in each image. The point spread function of the bright exoplanet host star is subtracted in all images, which are also high pass filtered.}\label{PICS}
\end{figure*}

For the calibration of our AstraLux imaging data we took dome- and sky-flats in the evening and/or morning twilight of each observing night. Darkframes were always taken directly before and in the same imaging setup, as used for the scientific observations. In each observing epoch several thousand short integrated images of $\gamma$\,Cep were taken with AstraLux with detector integration times ($DIT$), which range between 15 and 30\,ms. These images were reduced with our lucky-imaging pipeline, which performs the standard data-reduction (dark-subtraction, and flatfielding), followed by the measurement of the Strehl ratio reached in the individual images using the bright exoplanet host star as Strehl probe. The images were sorted by their Strehl ratio and only the 10\,\% of all images, which exhibit the highest Strehl ratios, were selected, registered, and combined. For the registration of the selected images, the position of the brightest pixel in the speckle pattern of the exoplanet host star was determined in all images, which were then shifted accordingly, and were eventually averaged to the final fully reduce image. Detail views on the reduced AstraLux images of all observing epochs, center on $\gamma$\,Cep\,A, are shown in Figure\,\ref{PICS}\hspace{-1.75mm}.

For the astrometrical calibration of the AstraLux detector we observed in each epoch the central region of the globular cluster M\,15 (center coordinates of the astrometric calibration field: $\text{RA(J2000)}=21^\text{h}29^\text{m}58^\text{s}$, $\text{Dec(J2000)}=+12^\circ09'56''$). At this position on the sky several dozens of stars are detected, for which an accurate astrometry is available in the early version of the third data release (EDR3) of the ESA-\textit{Gaia} mission \citep{gaiaedr3}. The used astrometric reference stars are homogenously distributed over the AstraLux field of view (24\,arcsec $\times$ 24\,arcsec), and exhibit G-band magnitudes in the range between 12.5 and 15.5\,mag, and a positional uncertainty of only 0.23\,mas, on average. In order to check the astrometrical stability of the instrument throughout an observing run in addition wide binaries (HIP\,59585, HIP\,72508, or HIP\,80953) were observed in the individual nights. The determined astrometrical calibration of the AstraLux detector is summarized for all observing epochs in Table\,\ref{TAB_ASTROCAL}\hspace{-1.75mm}.

\begin{table}[h!]
\caption{Astrometrical Calibration of the AstraLux detector. For each observing date ($ObsDate$) the determined pixel scale ($PS$) and detector position angle ($DPA$) are listed.}\label{TAB_ASTROCAL}
\centering
\begin{tabular}{ccc}
\hline
$ObsDate$      & $PS$            & $DPA$ \\
                & $[$mas/pixel$]$ & $[^\circ]$\\
\hline
21. Aug 2014 & $46.844 \pm 0.030$ & $-1.87 \pm 0.02$\\
27. Aug 2015 & $46.868 \pm 0.015$ & $-1.62 \pm 0.05$\\
26. Oct 2016 & $46.873 \pm 0.038$ & $-1.94 \pm 0.06$\\
30. Oct 2017 & $46.871 \pm 0.035$ & $-0.28 \pm 0.02$\\
05. Sep 2020 & $47.546 \pm 0.016$ & $-3.15 \pm 0.02$\\
\hline
\end{tabular}
\end{table}

The astrometry and photometry of the exoplanet host star were measured in the reduced AstraLux images, and those of the companion in the PSF subtracted images. The PSF subtraction was achieved via the roll-subtraction technique, as described by \cite{ginski2014}. With the software \verb"ESO-MIDAS" \citep{banse1983} we determined the positions of both objects in the AstraLux images via gaussian fitting, and their instrumental magnitudes with aperture photometry. The obtained astrometry of $\gamma$\,Cep\,B relative to the exoplanet host star is summarized for all observing epochs\linebreak in Table\,\ref{TAB_ASTROMETRY}\hspace{-1.75mm}. The orbital motion of the companion relative to its primary is clearly detected in our AstraLux images. Within about 6 years of epoch difference the angular separation of the companion significantly increased by $0.346\pm0.013$\,arcsec, while its position angle decreased by $11.5\pm0.5^\circ$. This is in good agreement with the expected variation of both parameters ($\Delta\rho = +0.345$\,arcsec, and $\Delta PA=-11.2\,^\circ$), derived with the orbital solution of the $\gamma$\,Cep binary system from \cite{neuhaeuser2007}.

\begin{table}[h!]
\caption{The AstraLux astrometry of the companion $\gamma$\,Cep\,B relative to its primary. For each observing date ($ObsDate$) the angular separation ($\rho$), and the position angle ($PA$) of the companion are listed.}\label{TAB_ASTROMETRY}
\centering
\begin{tabular}{ccc}
\hline
$ObsDate$   & $\rho$            & $PA$\\
$[$yr$]$   & $[$arcsec$]$      & $[^\circ]$\\
\hline
2014.6357    & $1.441 \pm 0.011$ & $226.05 \pm 0.41$\\
2015.6522    & $1.517 \pm 0.011$ & $223.89 \pm 0.41$\\
2016.8216    & $1.603 \pm 0.014$ & $220.32 \pm 0.49$\\
2017.8274    & $1.652 \pm 0.008$ & $219.70 \pm 0.29$\\
2020.6795    & $1.787 \pm 0.007$ & $214.60 \pm 0.23$\\
\hline
\end{tabular}
\end{table}

The magnitude difference between $\gamma$\,Cep\,B and its primary, as measured in all AstraLux images, is $\Delta \text{i}'=7.11\pm0.17$\,mag. With the apparent magnitude of $\gamma$\,Cep \citep[$\text{i}'=2.73\pm0.04$\,mag][]{ofek2008} this yields the apparent magnitude of the companion of $\text{i}'=9.84\pm0.17$\,mag. With the \textit{Gaia} EDR3 parallax of the exoplanet host star \citep[$\pi=72.517\pm0.147$\,mas][]{gaiaedr3} the absolute magnitude of $\gamma$\,Cep\,B can be determined, for which we obtain $M_{\text{i}'}=6.6 \pm 0.2$\,mag. At the age of the $\gamma$\,Cep system \citep[$6.6^{+2.6}_{-0.7}$\,Gyr][]{torres2007} this absolute magnitude is consistent with a low-mass stellar companion with a mass of $0.39 \pm 0.03$\,M$_\odot$, derived with the stellar evolutionary models from \cite{baraffe2015}. Thereby we applied the empirical color transformations from \cite{jordi2006}, to derive the absolute i$'$-band magnitudes of low-mass stars, whose properties are characterized by the used stellar evolutionary models. Furthermore, we select the 8\,Gyr isochrone of the models, which is the closest one to the given age of $\gamma$\,Cep. It should be mentioned that the derived mass estimate of the companion does not significantly vary for ages between 6 and 10\,Gyr, i.e. within the range of age adopted for the exoplanet host star, because the magnitude of low-mass stellar companions, such as $\gamma$\,Cep\,B, does not significantly change at ages of several Gyr. The mass of the companion, determined with our AstraLux photometry, is in good agreement with the mass of $\gamma$\,Cep\,B, derived from its Ks-band photometry $\text{Ks}=7.3\pm0.2$\,mag, determined by \cite{neuhaeuser2007}. Using the \textit{Gaia} EDR3 parallax of the exoplanet host star we obtain an absolute Ks-band magnitude of the companion of $M_{\text{Ks}}=6.60\pm0.20$\,mag, which yields a mass of $0.36 \pm 0.04$\,M$_\odot$. Hence, both the optical and near-infrared photometry of $\gamma$\,Cep\,B are consistent with a low-mass main-sequence star, which is located at the distance of the exoplanet host.

\begin{figure}
\resizebox{\hsize}{!}{\includegraphics{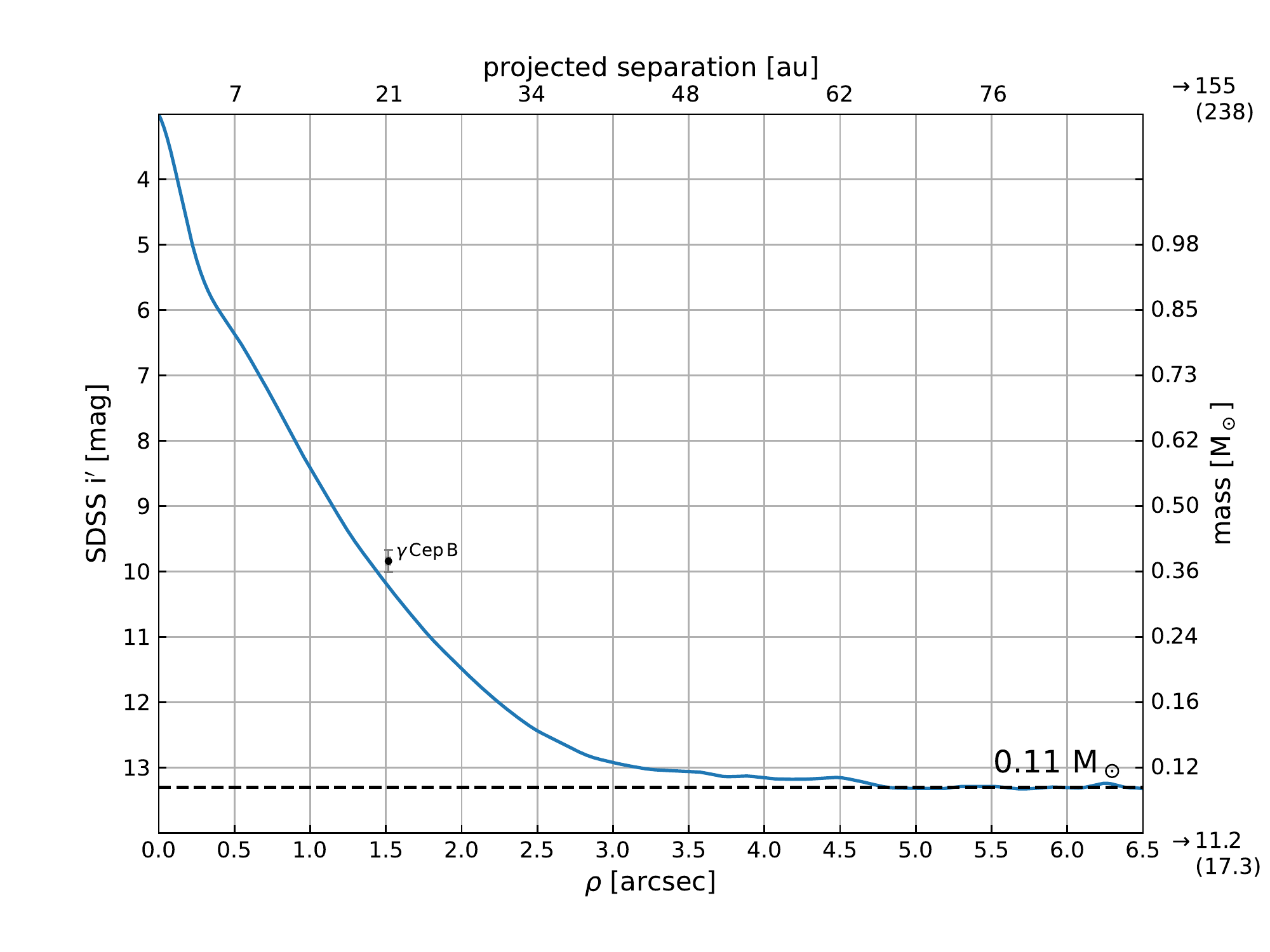}}\caption{The ($S/N=3$) detection limit, reached in our AstraLux i$'$-band images of $\gamma$\,Cep, plotted versus the angular separation (bottom axis), and projected separation (top axis) to the exoplanet host star, respectively. The maximal separation of companions, detectable in our AstraLux images, is given on the right top and bottom. The values in brackets are the maximal separations of detectable companions in the AstraLux images. The values without brackets are the radius of the field of view around the exoplanet host star, which is completely covered by our AstraLux observations.}\label{LIMIT}
\end{figure}

The detection limit reached in our AstraLux lucky-imaging observations is illustrated in Figure\,\ref{LIMIT}\hspace{-1.75mm}. Therein the projected separation is derived with the \textit{Gaia} EDR3 parallax of the exoplanet host star and the angular separation to the star, shown on the bottom axis of the diagram. The mass of detectable companions is derived, as described above, with the apparent magnitude (shown on the left axis of diagram), adopting again an age of 8\,Gyr. In the background limited region around the bright exoplanet host star beyond about 5\,arcsec (or 69\,au of projected separation) a limiting magnitude of $\text{i}'=13.3$\,mag is reached, which allows the detection of low-mass stellar companions down to $0.11$\,M$_\odot$.

Companions with angular separations up to 17.3\,arcsec (238\,au of projected separation) are detectable in our AstraLux images. The radial field of view, fully covered by the AstraLux observations, exhibits an angular radius of 11.2\,acsec, which allows the detection of companions with projected separations up to 155\,au. Beside $\gamma$\,Cep\,B no additional companion-candidate could be detected in our AstraLux images around the exoplanet host star. In particular, no other companions are found with projected separations smaller than about 73\,au, which is the closest long-term stable circumbinary orbit in the $\gamma$\,Cep binary system, calculated with the dynamical stability criterion from \cite{holman1999} using the system parameters ($a$, $e$ as given above, $mass(\text{A})=1.4\pm0.12$\,M$_\odot$, and $mass(\text{B})=0.409\pm0.018$\,M$_\odot$), determined by \cite{neuhaeuser2007}.

\section{FLECHAS Observations}

In addition to our lucky-imaging observations of $\gamma$\,Cep we also took follow-up spectra of the exoplanet host star, to measure the current RV of the star. We observed $\gamma$\,Cep in the course of the ongoing GSH Binary Survey, which is carried out with the \'Echelle spectrograph FLECHAS at the 0.9\,m telescope of the University Observatory Jena. The exoplanet host star was observed in three observing epochs between end of March 2020, and end of November 2021. In all observing nights three spectra of the star, each with an integration time of 150\,s, were taken in the 1$\times$1-binning mode of the FLECHAS detector, to reach the maximal resolving power of the instrument\linebreak ($R\sim9300$). Directly before the spectroscopy of $\gamma$\,Cep always three arcs (spectra of a ThAr-lamp), and three flats (spectra of a tungsten-lamp) were taken, each with an integration time of 5\,s for wavelength-, and flatfield-calibration, respectively. Furthermore, in each observing night three dark frames of all used detector integration times ($DIT$) were taken for dark-subtraction. Further details of the FLECHAS observations are summarized in the observing log\linebreak in Table\,\ref{TAB_FLECHAS_LOG}\hspace{-1.75mm}.

\begin{table}[h]
\caption{FLECHAS observing log. For each observing date ($ObsDate$) we list the number of taken spectra ($N_{\rm Specs}$), the detector integration time ($DIT$) of the individual spectra, the resulting total integration time ($TIT$) on target, the reached signal-to-noise ratio ($SNR$) in the fully reduced spectra at a wavelength of 6500\,\AA, as well as the average airmass ($X$) during the observations.}\label{TAB_FLECHAS_LOG}
\centering
\begin{tabular}{ccccccc}
\hline
$ObsDate$       & $N_{\rm Specs}$        & $DIT$    & $TIT$   & $SNR$ & $X$\\
                 &                    & $[$s$]$  & $[$s$]$ &    \\
\hline
25. Mar 2020     & 3                  & 150      & 450     & 640 & 1.58 \\
07. Sep 2020     & 3                  & 150      & 450     & 736 & 1.24 \\
20. Nov 2021     & 3                  & 150      & 450     & 696 & 1.14 \\
\hline
\end{tabular}
\end{table}

The data reduction was performed with the FLECHAS software pipeline \citep{mugrauer2014}, which includes darksubtraction, flatfielding, order extraction, wavelength calibration of the individual spectra, and their combination to the fully reduced spectrum of the star. The obtained FLECHAS spectra of $\gamma$\,Cep are all well exposed and exhibit on average a signal-to-noise ratio of $SNR=690$, at a wavelength of 6500\,\AA. We determined the RV of $\gamma$\,Cep with Doppler's law by measuring the wavelength of the cores of the most prominent Hydrogen lines ($H_{\alpha}$, $H_{\beta}$, and $H_{\gamma}$), detected in the FLECHAS spectra of the star, and comparing them with the laboratory wavelengths of the Balmer lines. The obtained RVs of $\gamma$\,Cep are summarized in Table\,\ref{RV}\hspace{-1.75mm}.

\begin{table}[h]
\caption{The RVs of $\gamma$\,Cep, obtained from our FLECHAS spectroscopy of the exoplanet host star.}\label{RV}
\centering
\begin{tabular}{cc}
\hline
$ObsDate$   & $RV$\\
$[$yr$]$   & $[$km\,s$^{-1}]$\\
\hline
2020.2329   & $-42.05 \pm 0.14$\\
2020.6871   & $-42.22 \pm 0.14$\\
2021.8873   & $-42.01 \pm 0.14$\\
\hline
\end{tabular}
\end{table}

The RV of $\gamma$\,Cep does not significantly change within the given epoch difference and we obtained an averaged RV of $-42.09\pm0.09$\,km\,s$^{-1}$, which is in good agreement with the expected RV of the exoplanet host star of $-42.02$\,km\,s$^{-1}$, as calculated with the orbit solution from \cite{neuhaeuser2007} for the mid-time of the FLECHAS spectroscopic observations.

\section{New Orbital Solution for the $\gamma$\,Cep Binary System}

As described in the last section, the RV measurements of $\gamma$\,Cep, obtained from our FLECHAS spectroscopy of the exoplanet host star, are in good agreement with the orbital solution of this binary system, presented by \cite{neuhaeuser2007}.

\begin{figure}[h!]
\resizebox{\hsize}{!}{\includegraphics{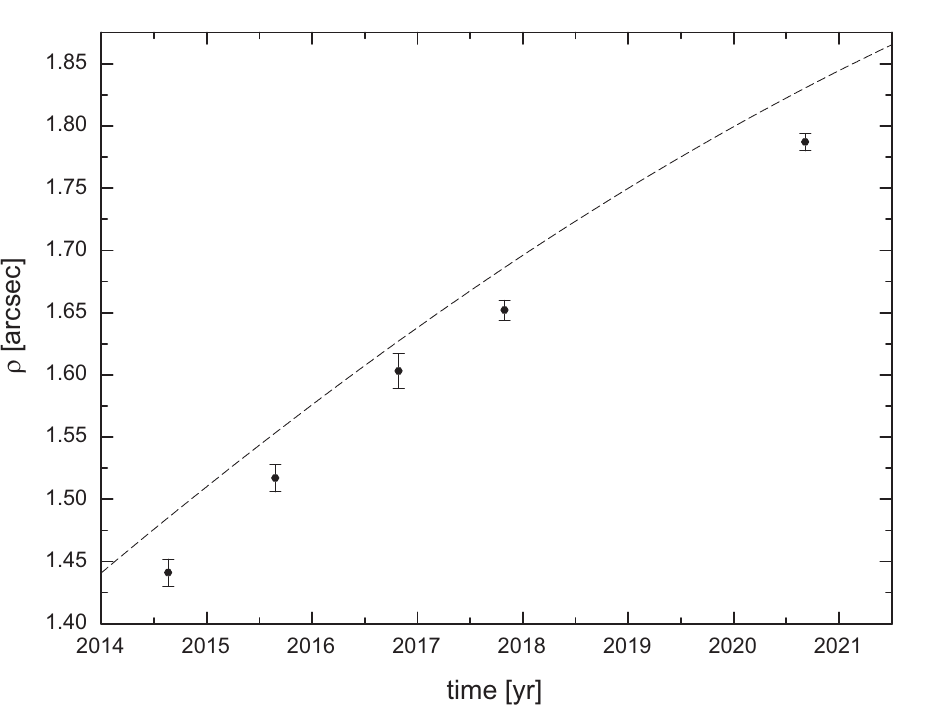}}\caption{The angular separation of $\gamma$\,Cep\,B to its primary, as measured in all AstraLux images, plotted versus time. The predicted angular separation of the companion according to the orbital solution from \cite{neuhaeuser2007} is shown as a black dashed line.}\label{RHO_TIME}
\end{figure}

In contrast, our AstraLux astrometry of $\gamma$\,Cep\,B, presented in section 2, significantly deviates from the orbital solution of \cite{neuhaeuser2007}. As shown in Figure\,\ref{RHO_TIME}\hspace{-1.75mm} for all AstraLux observing epochs the angular separation of $\gamma$\,Cep\,B to the exoplanet host star is smaller (on average by about 37\,mas) than predicted by the orbital solution from \cite{neuhaeuser2007}. This deviation of the measured and predicted angular separation of the companion is significant on the 3.9\,$\sigma$-level on average, clearly indicating that a redetermination of the orbital solution of the $\gamma$\,Cep binary system is necessary.

\begin{figure}
\resizebox{\hsize}{!}{\includegraphics{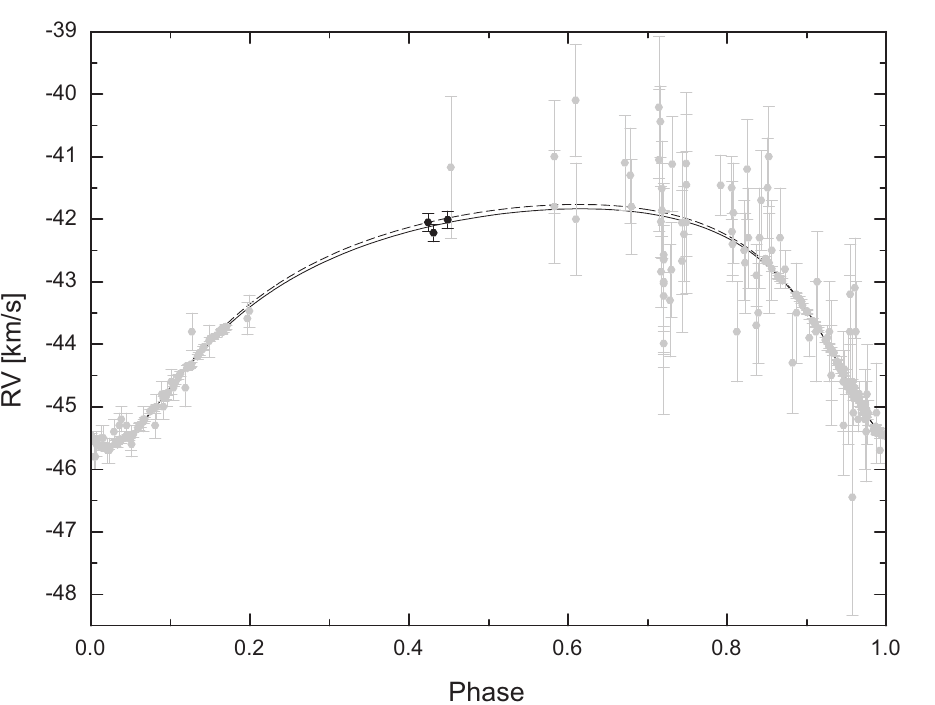}}\caption{The RV measurements of the exoplanet host star, plotted versus the orbital phase of the $\gamma$\,Cep binary system, together with the new orbital solution, presented here, and the one from \cite{neuhaeuser2007}, which are shown as black solid, and dashed line, respectively. The FLECHAS RVs are plotted as black filled circles and those from \cite{torres2007} as grey filled circles.}\label{ORBFIT_RV}
\end{figure}

For the calculation of a new orbital solution of the $\gamma$\,Cep binary system, we used the relative astrometry of $\gamma$\,Cep\,B from \cite{neuhaeuser2007}, as well as our AstraLux astrometry of the companion, i.e. 8 astrometric measurements in total. In addition, we included in the orbit calculation our new FLECHAS RVs, as well as those compiled by \cite{torres2007}. Hence, in total we have 310 RV measurements of $\gamma$\,Cep, which are illustrated in Figure\,\ref{ORBFIT_RV}\hspace{-1.75mm}. The spans of time, covered by the astrometric data of the companion relative to its primary and by the RVs of the exoplanet host star, which are used for the orbit determination of the $\gamma$\,Cep binary system, are 14 years (from 2006 to 2020) and 125 years (from 1896 to 2021) years, respectively.

By taking into account all astrometric and RV measurements inclusive their individual uncertainties we computed a combined orbital solution of the $\gamma$\,Cep binary system, using the code \verb"ORBITX" \citep{tokovinin1992}, which utilizes the Levenberg-Marquardt method to solve for all the Keplerian elements of the best fitting orbital solution. These elements are: the orbital period ($P$), the time of periastron passage ($T$), the semi-major axis ($a$) and eccentricity ($e$) of the orbit of the binary system, its inclination ($i$) to the plane of sky, the longitude of the ascending node ($\Omega$), the argument of periastron ($\omega$), the RV semi-amplitude ($K$) of the exoplanet host star, as well as the systematic RV ($\gamma$).

The Keplerian elements of the redetermined orbital solution of the $\gamma$\,Cep binary system are summarized in Table\,\ref{TAB_ORBFIT}\hspace{-1.75mm}. The orbit fit exhibits a reduced $\chi^2$-value of $\chi^2_\text{red}=1.0$, i.e. all astrometric and RV data agree well within their uncertainties with Keplerian motion, as expected.

\begin{table}[h]
\caption{Keplerian elements of the new orbital solution of the $\gamma$\,Cep binary system ($\chi^2_\text{red}=1.0$, based on 8 astrometric and 310 RV measurements), together with its derived parameters.}\label{TAB_ORBFIT}
\centering

\textbf{Keplerian Orbital Elements:}\vspace{1mm}\\

\begin{tabular}{rl}
\hline
$a[\text{arcsec}]=$           & \hspace{-3.5mm}$1.419 \pm 0.012$\\
$e=$                          & \hspace{-3.5mm}$0.4144 \pm 0.0066$\\
$i~[^\circ]=$                 & \hspace{-3.5mm}$120.18 \pm 0.27$\\
$\Omega~[^\circ]=$            & \hspace{-3.5mm}$18.32 \pm 0.78$\\
$\omega(\text{B})~[^\circ]=$  & \hspace{-3.5mm}$340.49 \pm 0.50$\\
$P~[\text{yr}]=$              & \hspace{-3.5mm}$66.84 \pm 1.32$\\
$T~[\text{yr}]=$              & \hspace{-3.5mm}$1991.581 \pm 0.048$\\
$K~[\text{km\,s}^{-1}]=$      & \hspace{-3.5mm}$1.898 \pm 0.014$\\
$\gamma~[\text{km\,s}^{-1}]=$ & \hspace{-3.5mm}$-42.989 \pm 0.027$\\
\hline
\end{tabular}

\vspace{3mm}\textbf{Derived System Parameters:}\vspace{1mm}\\

\begin{tabular}{rl}
\hline
$a~[\text{au}]=$                           & \hspace{-3.5mm}$19.56 \pm 0.18$\\
$f(\text{mass})~[\text{M}_\odot] =$        & \hspace{-3.5mm}$0.01303 \pm 0.00041$\\
$\text{mass}(\text{A})~[\text{M}_\odot] =$ & \hspace{-3.5mm}$1.294 \pm 0.081$\\
$\text{mass}(\text{B})~[\text{M}_\odot] =$ & \hspace{-3.5mm}$0.384 \pm 0.013$\\
\hline
\end{tabular}
\end{table}

\begin{figure}[h]
\includegraphics[width=\columnwidth, height=6.1cm]{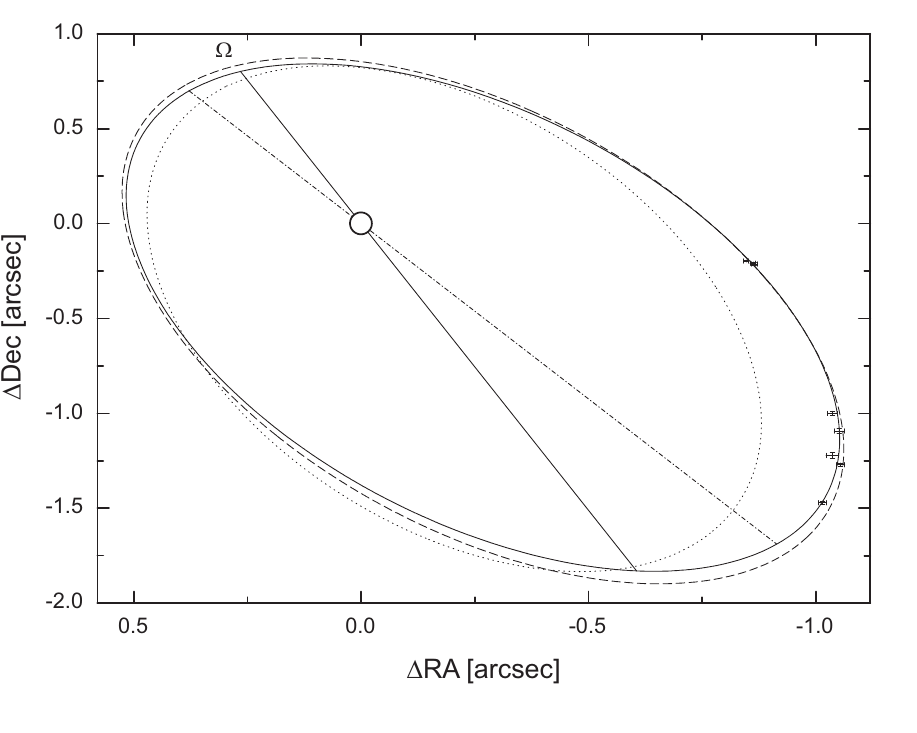}
\resizebox{\hsize}{!}{\includegraphics{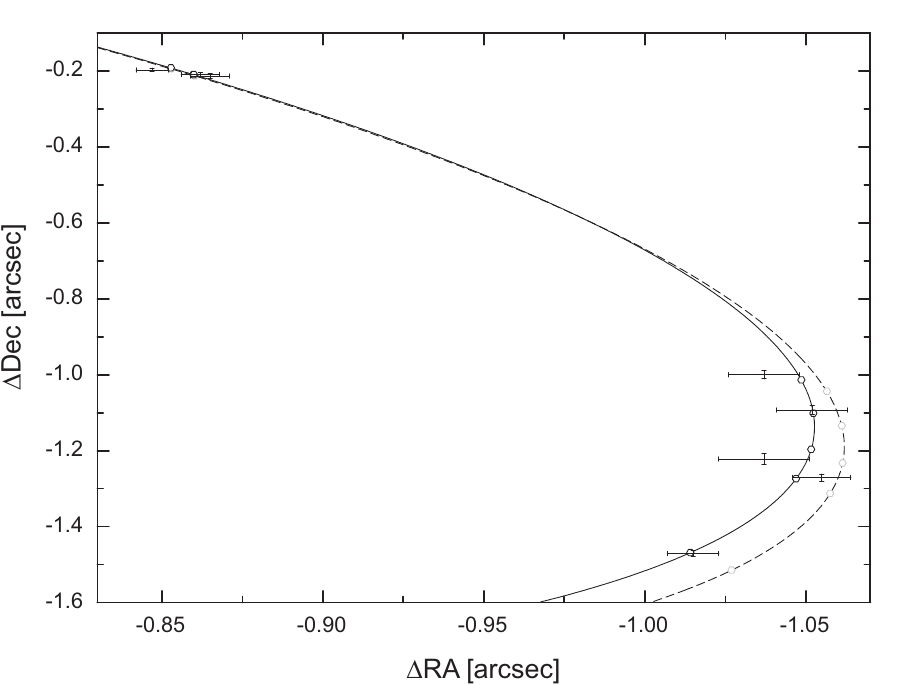}}
\caption{\textbf{Top:} The new orbital solution of the $\gamma$\,Cep binary system (black solid line) together with the solutions from \cite{torres2007} and \cite{neuhaeuser2007}, which are shown as a black dotted and dashed line, respectively. All astrometric measurements of $\gamma$\,Cep\,B, used to determine the new orbital solution, are plotted with their positional uncertainties. The apside line of the binary orbit is illustrated as a black dash-dotted line and its line of nodes with the ascending node $\Omega$ labelled, as black solid line. \textbf{Bottom:} Detail view on the astrometric data of $\gamma$\,Cep\,B and the orbital solutions, plotted in the same way as in the top pannel. The astrometric measurements from \cite{neuhaeuser2007} are located in the top left corner, our five AstraLux measurements in the lower right corner of the diagram, respectively. The black and grey circles on the orbits indicate the expected positions of the companion at the individual observing dates.}\label{ORBFIT_ASTRO}
\end{figure}

The new orbital solution of the $\gamma$\,Cep binary system is plotted together with all  astrometric and RV measurements, used for the orbit fitting, in Figure\,\ref{ORBFIT_ASTRO}\hspace{-1.75mm} and Figure\,\ref{ORBFIT_RV}\hspace{-1.75mm}, respectively.

We used the determined Keplerian elements of the best fitting orbital solution to derive the parameters of the $\gamma$\,Cep binary system, which are summarized in Table\,\ref{TAB_ORBFIT}\hspace{-1.75mm}. The semi-major axis of the binary system is determined with its apparent semi-major axis and the \textit{Gaia} EDR3 parallax\footnote{The Gaia EDR3 parallax of $\gamma$\,Cep ($\pi=72.517\pm0.147$\,mas) agrees well with previous parallax determinations of the star, e.g. $\pi=72.5\pm0.52$\,mas \citep{perryman1997}, $\pi=72.05\pm0.50$\,mas \citep{kharchenko2001}, or $\pi=72.69\pm0.41$\,mas \citep{neuhaeuser2007}, but is more accurate than these. Nevertheless, the astrometric six-parameter solution of $\gamma$\,Cep, listed in the \textit{Gaia} EDR3, exhibits a renormalized unit weight error (RUWE) of 3.212, indicating the multiplicity of the star in its astrometry, although it is not resolved by \textit{Gaia}.
We should note that $\gamma$\,Cep also has $G<6$\,mag, a brightness limit below which special and experimental position extraction is required. This could partially explain the higher RUWE. Furthermore, the given astrometric solution is not definite and exhibits a significant astrometric excess noise of 1.235\,mas. If we conservatively consider this noise as an additional uncertainty of the parallax of the star, this increases the uncertainties of the derived system parameters. In this case the uncertainties of the dynamical mass of the companion, of its primary, as well as of the semi-major axis of the $\gamma$\,Cep binary system, increase by factors of 1.5, 1.6, and 2.3, respectively. In contrast, the uncertainty in the photometric mass estimate of the companion is dominated by the photometric uncertainty and is therefore not appreciably altered by the assumed increased parallax uncertainty of the exoplanet host star.} of the exoplanet host star. The mass-function $f(\text{mass})$ of the system is calculated with its eccentricity and orbital period, as well as with the RV semi-amplitude of the exoplanet host star. The inclination and mass-function of the binary system yield the dynamical masses of its components. The derived dynamical mass of $\gamma$\,Cep\,A ($1.294\pm0.081$\,M$_\odot$) is in good agreement with the mass of the star $1.18\pm0.11$\,M$_\odot$, determined by \cite{torres2007}, using stellar evolutionary models and the determined absolute V-band magnitude and effective temperature of the exoplanet host star. The dynamical mass of $\gamma$\,Cep\,B ($0.384\pm0.013$\,M$_\odot$) is consistent with our mass estimates of the companion, derived with its i$'$- and Ks-band photometry, $0.39\pm0.03$\,M$_\odot$, and $0.36\pm0.04$\,M$_\odot$, respectively.
According to the new orbital solution of the $\gamma$\,Cep binary system, presented in this paper, $\gamma$\,Cep\,B will reach its apocenter at the end of 2024 and its angular separation to the exoplanet host star will increase until early 2029, when it will exhibit an angular separation of about 1.97\,arcsec. In particular the companion will be separated from its primary by more than 1.4\,arcsec until 2043 and hence remain observable by then with telescopes of the 2\,m-class or slightly below, using diffraction limited or Seeing enhanced imaging techniques, as it was demonstrated with our AstraLux observations, presented here. The components of the $\gamma$\,Cep binary system will pass through their next pericenters at mid of 2058. About three and a half years before that the exoplanet host star will reach its strongest acceleration in the radial direction of about $-0.36$\,kms$^{-1}$yr$^{-1}$. In the years before or after that date RV monitoring is most suitable to follow the change of the RV of $\gamma$\,Cep. In particular high precision RV data taken from now until 2048 (Phase$\sim$0.85), which cover the range of orbital phase where currently no accurate RVs are available, are very useful for further constraining the orbital solution of the $\gamma$\,Cep binary system. The same holds for astrometric observations to be carried out in the upcoming years, because the astrometric orbit of $\gamma$\,Cep\,B is covered so far only with a few data points. Therefore, during the next years we plan further follow-up high contrast AO, lucky-imaging, as well as spectroscopic observations of the $\gamma$\,Cep binary system.
\bibliography{mugrauer}
\section*{Acknowledgments}
We would like to thank the members of the technical staff of the Calar Alto observatory in Spain for all their help with the observations. In particular we thank Martin Seeliger, and Ana Guijarro, who carried out some of the AstraLux observations. MF acknowledges financial support from grant PID2019-109522GB-C5X/AEI/10.13039/501100011033 of the Spanish Ministry of Science and Innovation (MICINN) and from the State Agency for Research of the Spanish MCIU through the Center of Excellence Severo Ochoa award to the Instituto de Astrof\'{\i}sica de Andaluc\'{\i}a (SEV-2017-0709). We made use of data from the \verb"Simbad" and \verb"VizieR" databases, operated at CDS in France, and from the European Space Agency (ESA) mission \textit{Gaia} (\url{https://www.cosmos.esa.int/gaia}), processed by the \textit{Gaia} Data Processing and Analysis Consortium (DPAC, \url{https://www.cosmos.esa.int/web/gaia/dpac/consortium}). Funding for the DPAC has been provided by national institutions, in particular the institutions participating in the \textit{Gaia} Multilateral Agreement.
\vspace{-3mm}
\end{document}